\documentclass[]{spie}  

\usepackage{amsmath,amsfonts,amssymb}
\usepackage{graphicx}
\usepackage[colorlinks=true, allcolors=blue]{hyperref}
\usepackage{gensymb}
\usepackage{placeins}

\title{Designs for next generation CMB survey strategies from Chile}

\author[a]{Jason R. Stevens}
\author[b]{Neil Goeckner-Wald}
\author[c]{Reijo Keskitalo}
\author[d]{Nialh McCallum}

\author[b]{Aamir Ali}
\author[c,e]{Julian Borrill}
\author[d]{Michael L. Brown}
\author[b]{Yuji Chinone}
\author[a]{Patricio A. Gallardo}
\author[f,g]{Akito Kusaka}
\author[b,f,h]{Adrian T. Lee}
\author[i]{Jeff McMahon}
\author[a]{Michael D. Niemack}
\author[j]{Lyman Page}
\author[k]{Giuseppe Puglisi}
\author[l]{Maria Salatino}
\author[m]{Suet Ying D. Mak}
\author[n]{Grant Teply}
\author[d]{Daniel B. Thomas}
\author[a]{Eve M. Vavagiakis}
\author[o]{Edward J. Wollack}
\author[p]{Zhilei Xu}
\author[p]{Ningfeng Zhu}

\affil[a]{Department of Physics, Cornell University, Ithaca, NY, USA}
\affil[b]{Department of Physics, University of California, Berkeley, CA, USA}
\affil[c]{Computational Cosmology Center, Lawrence Berkeley National Laboratory, Berkeley, CA, USA}
\affil[d]{Jodrell Bank Centre for Astrophysics, University of Manchester, Manchester, UK}
\affil[e]{Space Sciences Laboratory, University of California at Berkeley, Berkeley, CA, USA}
\affil[f]{Physics Division, Lawrence Berkeley National Laboratory, Berkeley, USA}
\affil[g]{Department of Physics, The University of Tokyo, Tokyo, Japan}
\affil[h]{Radio Astronomy Laboratory, University of California, Berkeley, Berkeley, CA, USA}
\affil[i]{Department of Physics, University of Michigan, Ann Arbor, USA}
\affil[j]{Department of Physics, Princeton University, Princeton, NJ, USA}
\affil[k]{Department of Physics, Stanford University, Stanford, California, CA, USA}
\affil[l]{AstroParticle and Cosmology (APC) laboratory, Paris Diderot University, Paris, France}
\affil[m]{Imperial College London, London, UK}
\affil[n]{Department of Physics, UCSD, La Jolla, USA}
\affil[o]{NASA/Goddard Space Flight Center, Greenbelt, MD, USA}
\affil[p]{Department of Physics \& Astronomy, University of Pennsylvania, Philadelphia, Pennsylvania, PA, USA}

\authorinfo{Further author information: (Send correspondence to J.R.S.)\\J.R.S.: E-mail: jrs584@cornell.edu}

\begin{document} 
\maketitle

\begin{abstract}

New telescopes are being built to measure the Cosmic Microwave Background (CMB) with unprecedented sensitivity, including Simons Observatory (SO), CCAT-prime, the BICEP Array, SPT-3G, and CMB Stage-4.  We present observing strategies for telescopes located in Chile that are informed by the tools used to develop recent Atacama Cosmology Telescope (ACT) and \textsc{Polarbear} surveys. As with ACT and Polarbear, these strategies are composed of scans that sweep in azimuth at constant elevation.

We explore observing strategies for both small (0.42 m) aperture telescopes (SAT) and a large (6 m) aperture telescope (LAT). We study strategies focused on small sky areas to search for inflationary gravitational waves as well as strategies spanning roughly half the low-foreground sky to constrain the effective number of relativistic species and measure the sum of neutrino masses via the gravitational lensing signal due to large scale structure. We present these strategies specifically considering the telescope hardware and science goals of the SO, located at $23\degree$ South latitude, $67.8\degree$ West longitude.

Observations close to the Sun and the Moon can introduce additional systematics by applying additional power to the instrument through telescope sidelobes. Significant side lobe contamination in the data can occur even at tens of degrees or more from bright sources. Therefore, we present several strategies that implement Sun and Moon avoidance constraints into the telescope scheduling.

Scan strategies can also be a powerful tool to diagnose and mitigate instrumental systematics either by using multiple scans to average down systematics or by providing null tests to diagnose problems. We discuss methods for quantifying the ability of an observation strategy to achieve this. 

Strategies for resolving conflicts between simultaneously visible fields are discussed. We focus on maximizing telescope time spent on science observations. It will also be necessary to schedule calibration measurements, however that is beyond the scope of this work. The outputs of this study are algorithms that can generate specific schedule commands for the Simons Observatory instruments.

\end{abstract}

\keywords{Observing Strategy, Simons Observatory, Cosmic Microwave Background}

\section{INTRODUCTION}
\label{sec:intro}  


Precision measurements of the anisotropy of the cosmic microwave background (CMB) have become a cornerstone of modern cosmology. These anisotropies have both temperature and polarization components. The polarization anisotropies can be decomposed into both even-parity ($E$-mode) and odd parity ($B$-mode) signals. Primordial scalar (density) fluctuations can only source $E$-mode polarization. $B$-mode polarization in the CMB can be created by the gravitational lensing of the primordial $E$-mode polarization or by primordial tensor  perturbations (gravitational waves). The lensing $B$-mode signal is expected to peak at sub-degree scales while the gravitational wave signal is expected to peak at degree scales. A detection of the gravitational wave signal would provide powerful evidence for the theory of cosmic inflation. Precision measurements of the lensing signal through the $B$-mode channel or four point statistics could detect a non-zero sum of neutrino masses or additional relativistic species in the universe beyond what is predicted by the standard model.\cite{2016arXiv161002743A}

Current and next generation CMB telescopes are seeking to make increasingly sensitive maps of the the CMB polarization. The observation strategy used by an instrument can have a major impact on the ability to resolve the polarization anisotropies. The Simons Observatory (SO) will deploy a 6 meter crossed Dragone telescope\cite{Parshley2018} designed to measure the lensing $B$-mode signal and three 0.42 m refractive telescopes to search for the primordial gravitational wave signal\cite{Galitzki2018}. Each class of telescope will observe the sky following a strategy optimized for its portion of the SO science goals. The angular scale or multipole $\ell \approx 360 / \theta ^\circ$ range targeted by an instrument is a primary consideration.

\begin{figure}
\begin{center}
\begin{tabular}{l r}
\includegraphics[height=5cm]{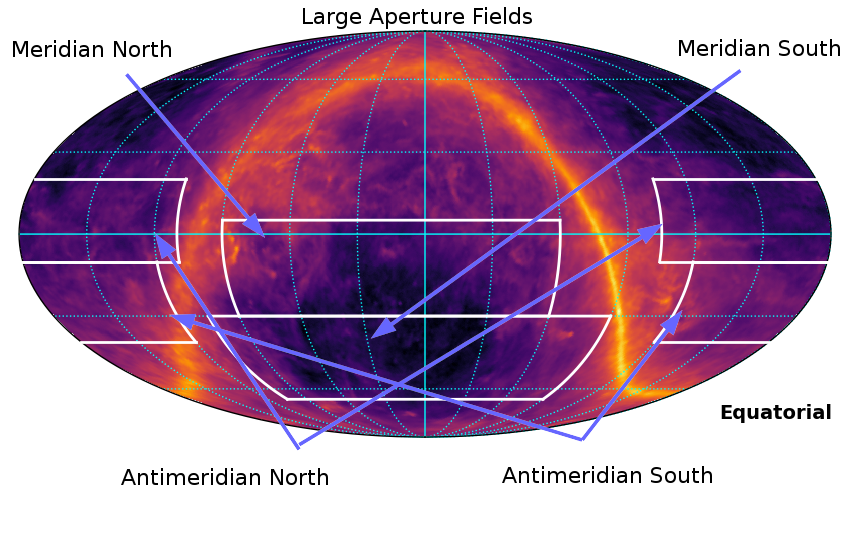} &
\includegraphics[height=5cm]{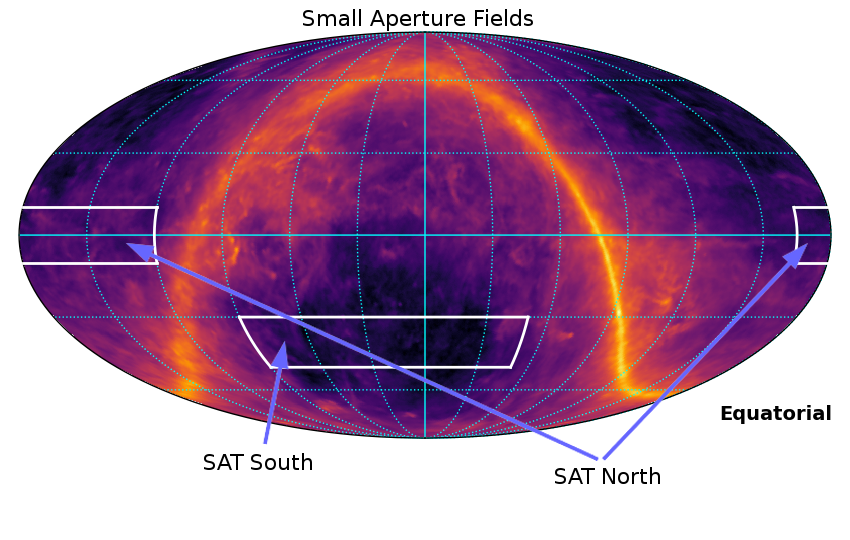}
\end{tabular}
\end{center}
\caption{
\label{fig:lat3}
Left: Fields for a proposed large aperture, large area survey over the Planck dust intensity map. The fields inside the boxes cover 17{,}095 square degrees and represent the region targeted by the telescope boresight. These are the fields used by the Classical strategy. Right: Fields for a proposed small aperture survey. The boresight targets 4{,}920 square degrees, but in practice the observed area is much larger due to the $35\degree$ field of view of the telescopes.\cite{Galitzki2018} These fields are used by both the Classical and Opportunistic strategies.}
\end{figure}


Measuring the reionization bump at very low multipoles ($\ell \lesssim 10$) requires a dedicated scan strategy \cite{2014SPIE.9153E..1IE,2014JLTP..176..733M,2016JLTP..184..786O} and will not be a target of the SO. The bulk of the inflationary $B$-mode signal will be at the recombination bump around $50 < \ell < 200$ and will be the primary target of the SO small aperture telescopes. The lensing $B$-mode signals will be a primary target of the SO large aperture telescope and peaks around $500 < \ell < 2000$

Typically, scan strategies for instruments searching for the inflationary $B$-mode feature have observed a small area ($f_{\rm sky}\sim 1\%$) in the lowest foreground regions of the sky.\cite{2015PhRvL.114j1301B,2010SPIE.7741E..1NF,2010arXiv1008.3915E,2013ApJ...768....9B} Since the $B$-mode polarization due to tensor perturbations is expected to be small, prior to a B-mode detection it can be advantageous to scan a small portion of the sky to concentrate the instrumental sensitivity into a small area and reduce noise variance in the final map as much as possible.

Targeting large angular scale polarization also necessitates having a strategy composed of scans with large throws at the same location in ground coordinates, which is necessary due to the statistical properties of detector noise. These scans should also have a long duration to allow the sky to move enough relative to the ground to orthogonalize signals fixed in the sky and signals fixed on the ground. There is an essential trade off between these geometric considerations and the statistical benefits of scanning a smaller fraction of sky.

Telescopes designed to measure small scale modes in the CMB are driven by a separate set of considerations. The need for large throws and long scans are mitigated at smaller feature sizes. Small scale temperature, E-mode, and B-mode anisotropies are expected due to primordial effects, gravitational lensing, galaxy clusters, and other sources, so cross correlation (and therefore patch overlap) with optical surveys such as LSST and DESI are valuable. These science cases generally favor a larger sky area to reduce the sample variance due to measuring a finite number of modes on the sky. Since there is no similar tension between sky area and scan geometry, these scan strategies have a significantly more open parameter space.

At the map depths projected for CMB Stage-4 \cite{2016arXiv161002743A} the lensing signal will become a significant foreground for inflation measurements. This signal can be removed given a measurement of the lensing deflection field. This template can be measured using an external tracer of large scale structure \cite{2015PhRvD..92d3005S} or using the lensing signal measured from a high-resolution CMB instrument. For this reason it is advantageous to have significant area overlap between small and large aperture scanning strategies.

A significant systematic in CMB measurements is galactic foregrounds. These can be subtracted using information from multiple frequencies, but the subtraction becomes difficult if the foreground contamination is large. CMB surveys therefore target low-foreground regions off the galactic plane, as in Figure \ref{fig:lat3}. A scan strategy must additionally ensure good overlap between the regions scanned at different frequencies. Thus, for the LAT, it is preferred to cover fields significantly larger than the field of view, which would minimize the area that different frequencies do not overlap compared to the total coverage.

Additionally, there is a natural complementarity in the observation strategies between experiments in the Atacama and the South Pole. Instruments in the Atacama have access to a larger sky area and can achieve closer to isotropic mode coverage on the sky by observing the same point on the sky at multiple elevations. However, Chilean instruments must contend with a less stable and uniform atmosphere, less stable ground pickup due to lack of a featureless horizon, as well as scheduling constraints due to patches setting below the horizon. A detailed comparison of the trade offs between the two sites is beyond the scope of this paper. 

We explore techniques for developing observing strategies for Chilean telescopes with a particular emphasis on application to SO. Section \ref{sec:fieldsandstrategies} describes two different approaches to generating observing strategies: ``classical'' and ``opportunistic;'' Section \ref{sec:sunandmoonavoidance} describes how these strategies avoid observing close to the Sun and the Moon; Section \ref{sec:proposedstrategies} describes examples of both classical and opportunistic observing strategies for the Simons Observatory's Large Aperture Telescope and Small Aperture Telescopes; Section \ref{sec:comparison} provides a few direct comparisons between the two styles of strategies and proposes some figures of merit for evaluating them.

\section{FIELD AND STRATEGY CONSIDERATIONS}
\label{sec:fieldsandstrategies}

We have developed two different approaches to creating observing strategies for these telescopes. A ``classical'' CMB strategy is similar to strategies on current CMB telescopes and relies on switching between a largely predetermined set of patches and scans. The ``opportunistic'' strategies use a new set of software that breaks the desired fields into tiles and prioritizes them. The strategy then chooses tiles to observe on the fly based on priority, accumulated depth on each tile and Sun and Moon avoidance. We will compare these methods throughout later sections.

\subsection{Creating and Optimizing Classical CMB Strategies}

Here, by ``classical CMB strategy'' we mean one developed with a similar approach to previous CMB telescopes including Advanced ACTPol\cite{debernardis2016}. First, target fields are selected. A contiguous area may be split into multiple fields (for instance, a north and a south field) to reduce either the scan throw or the required observation length. Each field is observed at multiple elevations as an instrumental systematic mitigation strategy, as discussed in Section \ref{subsec:systematics}.

\begin{figure} [ht]
\begin{center}
\begin{tabular}{l r} 
\includegraphics[height=6cm]{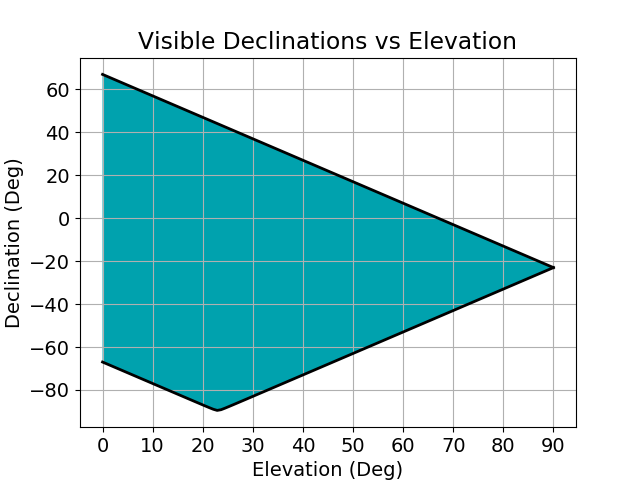} &
\includegraphics[height=6cm]{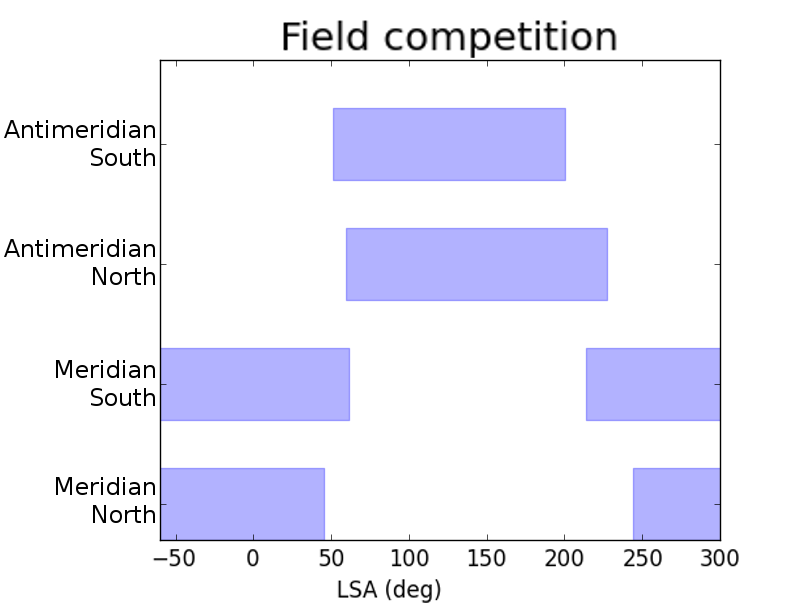}
\end{tabular}
\end{center}
\caption[example] 
{ \label{fig:decaccess} 
Left: Declinations visible from the Atacama desert, at $23 \degree$ south latitude, as a function of telescope scan elevation. The shaded region is visible. Right: Ranges of local sidereal angles for which the various fields in the proposed classical large aperture strategy are visible assuming a $40\degree$ elevation horizon. This minimum observing elevation can be tuned as necessary. Fields that overlap in local sidereal angle (LSA) can't be observed simultaneously. It is desirable for there to be some field visible for every LSA. These ranges are determined by the field definitions, scan elevation and whether a patch is rising or setting. }
\end{figure}

The range of declinations observable from the Atacama as a function of elevation is shown in Figure \ref{fig:decaccess}. Conflicts between fields are initially resolved by comparing the local sidereal angle of the fields at a given elevation or set of elevations to achieve an efficient schedule. The elevations and fields themselves can be adjusted to further resolve conflicts and improve efficiency. Several separate schedules are generated, each prioritizing observations at a specific elevation in the rising or the setting sky. If a large area has been split into multiple fields each field is observed in a separate schedule. One schedule is observed each day cycling through all of them periodically.

In the Advanced ACTPol nighttime strategy, there are northern and southern fields observed as they rise or set at three possible elevations ($40\degree$, $45\degree$, and $47.5\degree$). This makes for a total of $2~\textrm{(fields)} \times~ \allowbreak 2~\textrm{(rising, setting)} \times~ \allowbreak 3~\textrm{(elevations)} = 12$ night time schedules that are cycled through every 12 nights. Uniformly distributing time among each schedule helps maintain good cross linking. The Advanced ACTPol day time schedules cover smaller areas\cite{debernardis2016}.

The current Polarbear-1 scan strategy is a minor perturbation on this scheme. There is one deep field close to the South celestial pole. This field is scanned as it rises, transits and sets. There are a set of ten schedules changing the elevation of the transit scan that are cycled through daily.

\subsection{Opportunistic Scheduling Algorithm}

We have also developed a new approach to scheduling the observations that dynamically adjusts the priority of acquirable targets. The inputs to the scheduler are the target patches, their relative priorities, observing period and boundary conditions (allowed observing elevations, Sun and Moon avoidance radii). The scheduler then builds the observing schedule from a given start time, always observing the highest priority acquirable target. The relative priorities are dynamically modulated by the amount of already-scheduled integration time. Here is a breakdown of the scheduling algorithm:
\begin{enumerate}
\item \label{item:list} Collect a list of targets presently within the allowed observing elevation. For a target to be considered, it has to be \begin{itemize}
	\item rising and completely below maximum observing elevation or
	\item setting and completely above minimum observing elevation
\end{itemize} with some part of the target inside the allowed elevation range.
\item Sort the target list by relative priority. Priority is made of input relative priority, prior scheduled scanning time and other possible modifiers (such as observing elevation and required slew time to reach the target).
\item \label{item:available} If no targets are available, advance the scheduler time by a fixed step and return to step \ref{item:list}. Note that this represents lost observation time and is avoided as much as possible.
\item Attempt scheduling a constant elevation scan over the highest priority target.  The scan will fail if the Sun or Moon is found too close to the target\label{item:attempt}.
\item If a scan is successfully scheduled, advance the scheduler time and return to step \ref{item:list}. If the scan failed, remove the highest priority item from the target list and return to step \ref{item:available}.
\end{enumerate}

Internally, the targets are represented as sets of control points called ``corners.'' The user has a choice of defining the targets as ranges in right ascension and declination, disc center and radius or as corners of a polygon. Our implementation uses PyEphem\footnote{\url{http://rhodesmill.org/pyephem}. PyEphem back end comes from XEphem (\url{http://www.clearskyinstitute.com/xephem})} to carry out the time-dependent coordinate transformations between the celestial and horizontal coordinate systems. PyEphem is also used to predict the positions of the Sun and the Moon.

The input priorities for the targets may either be quantized into primary and secondary targets or they may reflect the expected level of foreground emission in the targets. Nested and overlapping targets are also supported. Our implementation includes options to alternate between rising and setting scans as well as cycling through different observing elevations.


The algorithm works best when, at any given time, multiple targets can be acquired and assigned relative priorities. To this end, it is natural to break down large patches into smaller, possibly overlapping tiles. The optimal shape and overlap of such tiles must be optimized for the instrument and science targets. In particular, an instrument targeting the inflationary $B$-mode signal will opt for larger tiles to improve sensitivity to large angular scales.


It is conceivable to run the scheduler in real time, assigning actual noise and yield numbers to past scans to modulate the relative priorities.

\section{Sun AND Moon AVOIDANCE}
\label{sec:sunandmoonavoidance}

\subsection{Classical Strategy Sun and Moon Avoidance}
\label{sec:avoidance}

For the classical strategies, we implement Sun and Moon avoidance by taking advantage of the distinct daily schedules. The schedules are put into pairs of ``complementary'' schedules. Rising and setting scans of different fields at different elevations are mixed and matched so that complements point the telescope in different directions. This is done to maximize the probability that a complementary patch will be available if a scan is blocked by the Sun or Moon. Since the strategies are in pairs, each strategy still gets approximately the same amount of time it would if avoidance were not implemented. Also, the fact that the Sun (and Moon) spends approximately equal time at each right ascension over the year aids in preventing one part of the sky from accumulating or losing much observation depth. Examples of chosen complements for our classical strategies are described in tables \ref{tab:lat_complements} and \ref{tab:sat_complements}.

When the Sun or Moon passes within an exclusion radius of the telescope boresight, the telescope is switched to the complement strategy. This method is used to avoid the Sun during the day (defined as any time the Sun is above the horizon) and to avoid the Moon at night. Moon avoidance is not applied during the day because Sun avoidance takes priority and it is difficult to simultaneously avoid both objects with this method. 

Under some circumstances, the Sun or Moon can be in the exclusion region for both a strategy and its complement. There is an additional cut applied to the scheduling when the complement approach fails. In practice this cut does not significantly impact the observing efficiency. In the final simulations neither the Sun nor the Moon are ever within the exclusion region.

This method had been used with a $35\degree$ exclusion radius on the Advanced ACTPol night strategy to avoid the Moon since the 2017 season. In the POLARBEAR-1 large patch observation the field location is far enough below the ecliptic plane that Sun avoidance is not an issue and no such complement strategy is necessary.

\subsection{Opportunistic Scheduler}

The opportunistic scheduler will not schedule scans that cause the boresight to enter the avoidance radius around the Sun or the Moon. The scheduler recovers from Sun avoidance by scheduling more integration time in targets that have been blocked by the Sun earlier. One useful extension of the tiling approach is to make the full patch a primary target and the comprising tiles secondary. Individual tiles are targeted only when the Sun prevents conducting full patch scans. This is useful from a low-$\ell$ perspective since longer scans provide better sensitivity to large angular scale modes. Through the season different tiles become acquirable and the scheduler balances their individual integration times based on their relative priorities.

\section{Proposed Strategies}
\label{sec:proposedstrategies}

Here we will discuss proposed scan strategies for the small and large aperture telescopes produced using both the classical and opportunistic scheduling approach. The strategies are simulated for one calendar year. It should be noted that this does not include telescope down time due to maintenance or the Altiplanic winter.

\subsection{Large Area Strategies for Large Aperture Telescopes}

Chilean CMB surveys designed to look at very large areas are constrained by several factors. First, the Chilean sky has a limited range of accessible declinations. Since the site in the Atacama desert is at $23\degree$ South latitude, the sky above a declination of $67\degree$ is never observable. There is a preference for observing at higher elevations to reduce atmospheric loading on the detectors, although, if the elevation is too high, the scan will become slow and short, leading to degraded sensitivity. It is preferable to avoid regions near the galactic plane since maps of these regions are heavily contaminated by foregrounds. The fact that these strategies target large areas makes achieving a high observation efficiency easier than it would be in a smaller area scan because there are more potential targets available at a given time.

\subsubsection{Classical Strategy}
\label{sec:lat3}
\begin{figure}
\begin{center}
\begin{tabular}{l r}
\includegraphics[height=5cm]{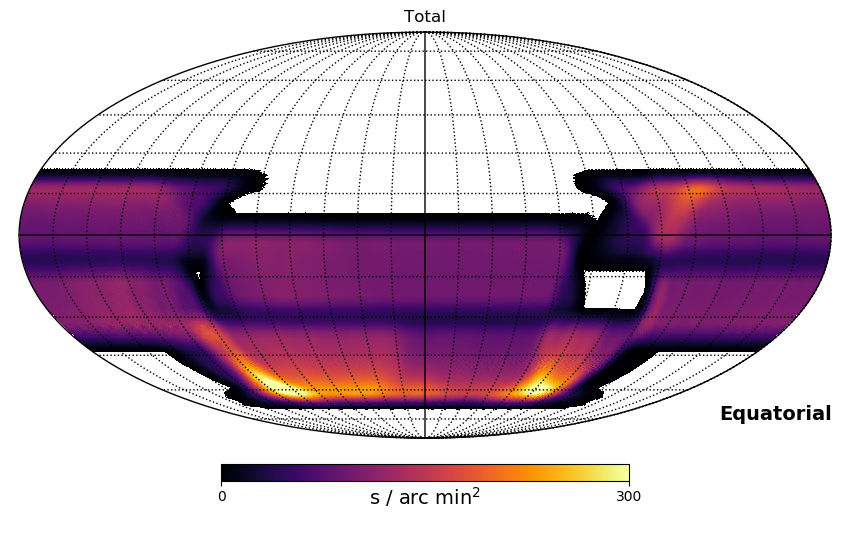} &
\includegraphics[height=5cm]{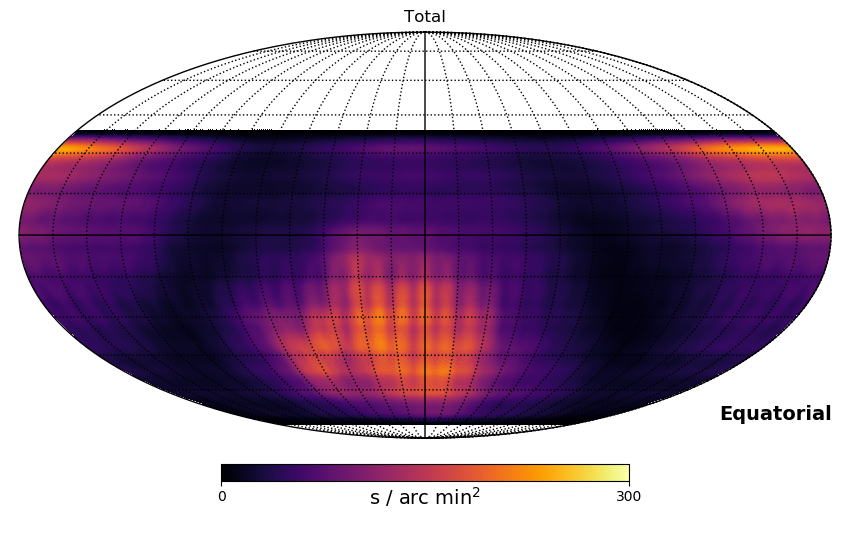} 
\end{tabular}
\end{center}
\caption{
	\label{fig:lat_comparison}
	Resulting integration depths from the classical (left) and opportunistic (right) schedules.
}
\end{figure}

The fields shown in Figure \ref{fig:lat3} enclose more than 17{,}000 square degrees or 41\% of the sky. The simulated observation depth map of such a strategy is shown in Figure \ref{fig:lat_comparison}. The field of view used in the simulations is $8\degree$. The full field of view passes outside of the target regions, so 58\% of the sky is actually observed, although, a significant amount of the total area is only observed at low depths, as quantified in Section \ref{sec:fsky}.

This strategy is based very closely on the Advanced ACT nighttime strategy. Each day, a strategy will prioritize observations of either the northern fields or the southern fields. Also, one of three elevations is chosen to observe at exclusively: $40\degree$, $44\degree$, or $47\degree$. Note that Figure \ref{fig:decaccess} shows that $47\degree$ is the highest elevation that the maximum declination of $20\degree$ can be observed. In other words, moving to a higher elevation scan would require sacrificing sky area at the northern edge of the patch. Choosing observations to match these twelve types help resolve conflicts between fields on a given day and assure that observation time of each part of the sky can be kept roughly even.

Since cycling through these twelve days in an equal proportion would favor the southern fields due to their declination range, this strategy attempts to observe the northern fields more often than the southern fields in a 3 to 2 ratio. This increases depth in the northern area and improves the evenness between fields.

During the day, Sun avoidance is applied as described in Section \ref{sec:avoidance}. The exclusion radius is $30\degree$. The complement pairs are listed in table \ref{tab:lat_complements}. After Sun and Moon avoidance are applied, any remaining time where the Sun or Moon is within $30\degree$ of the boresight is removed from the schedule. For this analysis we simulate 60 seconds of idle time between scans. This simulates the time it takes the telescope to move to its new position. The resulting strategy maintains a 91.3\% observing efficiency after all down time and cuts. It should be noted that this number does not include time to calibrate or tune the array and will be somewhat smaller in practice. These considerations are beyond the scope of this paper.

\begin{table}
\begin{center}
\begin{tabular}{| l | r |}
\hline
Strategy & Complement \\
\hline
Rising, $40\degree$ elevation, north fields & Setting, $47\degree$ elevation, south fields \\
Rising, $44\degree$ elevation, north fields & Setting, $44\degree$ elevation, south fields \\
Rising, $47\degree$ elevation, north fields & Setting, $40\degree$ elevation, south fields \\
Rising, $40\degree$ elevation, south fields & Setting, $47\degree$ elevation, north fields \\
Rising, $44\degree$ elevation, south fields & Setting, $44\degree$ elevation, north fields \\
Rising, $47\degree$ elevation, south fields & Setting, $40\degree$ elevation, north fields \\
\hline
\end{tabular}
\end{center}
\caption{
\label{tab:lat_complements}
Complement pairs for the large aperture, large area classical strategy. Each line represents one pair of strategies, each of which is designed to observe several fields from Figure \ref{fig:lat3} at a fixed elevation and exclusively either rising or setting. ``north'' or ``south'' here refer to which field is chosen to observe among a pair of fields at similar right ascensions; ``north'' referring the field that is closer to the north terrestrial pole (greater declination), and ``south'' referring to the field that is closer to the south terrestrial pole (lesser declination).}
\end{table}

\begin{figure}
\begin{center}
\begin{tabular}{l r}
\includegraphics[height=5cm]{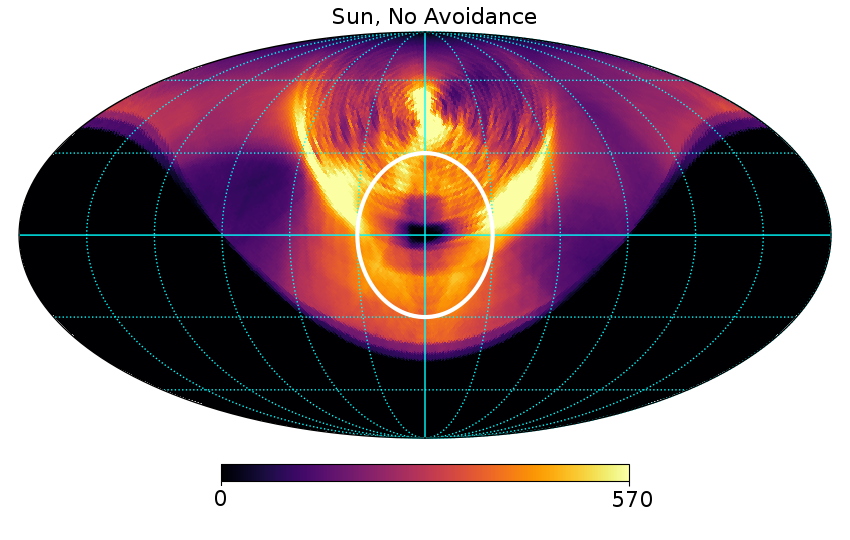} &
\includegraphics[height=5cm]{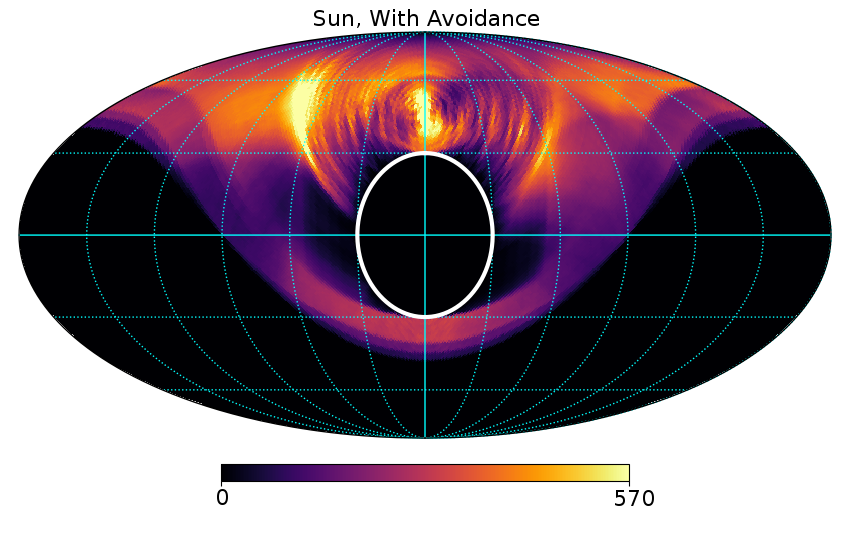} \\
\includegraphics[height=5cm]{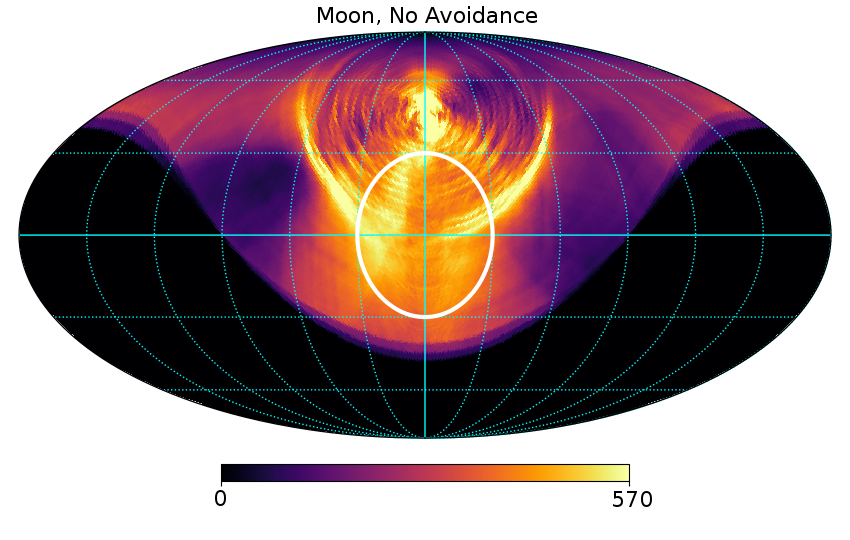} &
\includegraphics[height=5cm]{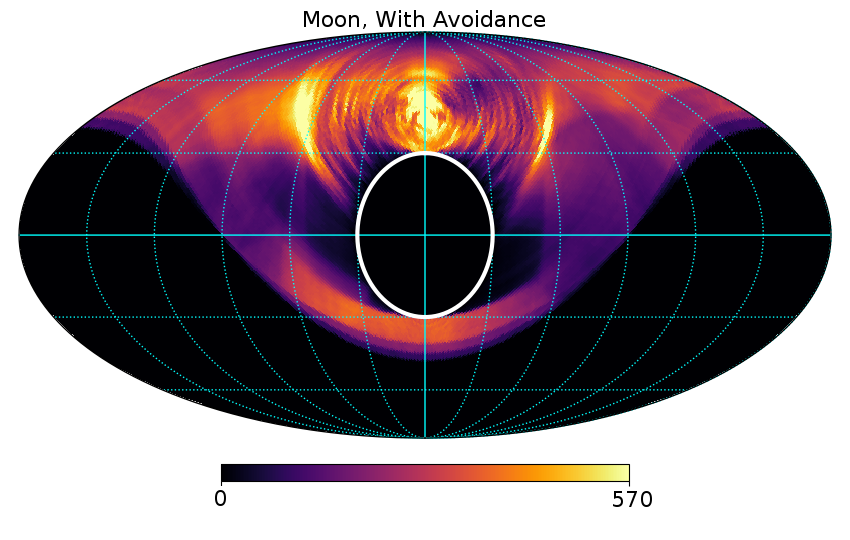}
\end{tabular}
\end{center}
\caption{
\label{fig:lat3Sunavoidance}
Sun (Moon) avoidance for the strategy described in Section \ref{sec:lat3}. The coordinate system is defined such that the boresight is at the center of the map and the polar axis of the coordinate system is perpendicular to the direction the telescope is pointing. The central meridian corresponds to directly above and below the boresight in local coordinates. The depth shows the amount of time the Sun (Moon) is in each position in the sky relative to the boresight. The white circle shows the $30\degree$ exclusion region. Left maps show the strategy with neither Sun nor Moon avoidance, and right maps show the strategy with Sun and Moon avoidance. Top maps are Sun maps, and bottom maps are Moon maps.
}
\end{figure}

\subsubsection{Opportunistic Strategy}

\begin{figure}
\begin{center}
\includegraphics[width=\textwidth]{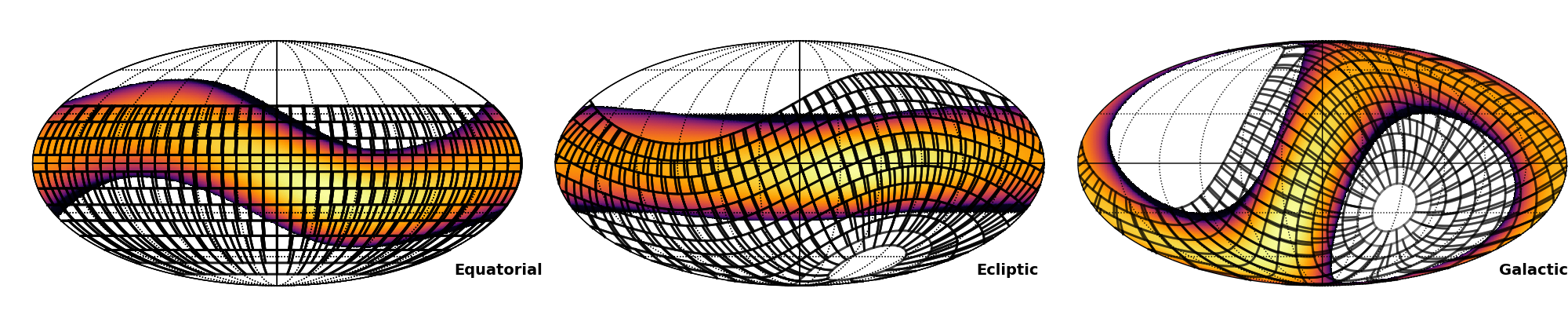}
\end{center}
\caption{
	\label{fig:lat_tiles}
	The tiling scheme for the opportunistic large area schedule. Each isolatitude band
    is made of uniform tiles whose width is modulated by latitude for approximately equal
    area. The colored band is formed of pixels that fall into the $30^\circ$ avoidance 
    region around the Sun or the Moon at any point during the season. The bands overlap by
    $5$ degrees in the North-South direction and are 10 degrees wide. The relative priority
    of a tile is determined by the dust intensity (Fig.~\ref{fig:lat3}).
}
\end{figure}

We created an opportunistic observing schedule by tiling the sky between $-75^\circ$ and $35^\circ$ of declination into rectangular $10^\circ \times 20^\circ$ (RA$\times$DEC) tiles with a $10^\circ$ overlap in the declination direction. The width in RA was adjusted as a function of DEC to make the tile area as uniform as possible while dividing each constant latitude $20^\circ$ band to an integer number of tiles. The tiling is shown in Figure~\ref{fig:lat_tiles}. Relative priority of the tiles was adjusted to reflect the square root of the expected dust intensity in each tile. The priority guides the scheduler to dedicate more integration time to cleaner parts of the sky and demonstrates how the opportunistic algorithm can be adjusted.
The allowed observing elevations were between $30^\circ$ and $70^\circ$. The scheduler was run for the entire calendar year 2019, producing an observing schedule with a 97.8\% observing efficiency while maintaining a $30^\circ$ avoidance region between the boresight and the Sun and the Moon.

\subsection{Small Aperture Strategies}
\begin{figure}
\begin{center}
\includegraphics[width=\textwidth]{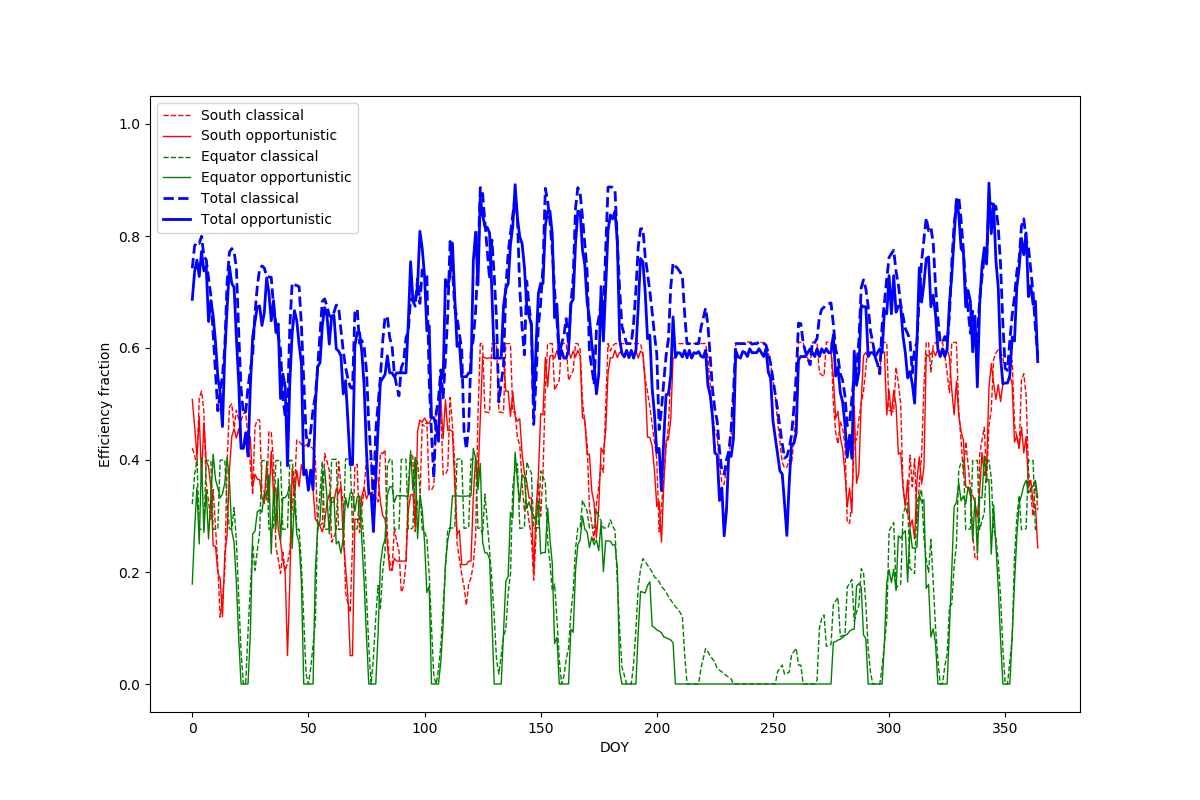}
\end{center}
\caption{
	\label{fig:sac_eff}
	Efficiency versus time of year for the SAT opportunistic and classical scan strategies. The monthly oscillation is due to the Moon avoidance criteria and the yearly oscillation is due to the Sun avoidance. 
}
\end{figure}

These strategies are specifically designed with the hardware of the SO Small Aperture Telescope in mind. This telescope will have a $35\degree$ field of view, and its scan elevations will be limited to the range $50\degree$ to $70\degree$. Due to the larger field of view and comoving baffle design considerations, both the classical and opportunistic strategies here avoid the Sun and Moon coming within $45\degree$ of the boresight. They also both limit themselves to the same observing regions, depicted in Figure \ref{fig:lat3}. This should enable a direct comparison of the classical and opportunistic strategies.
Since the observed region is smaller, the telescope may at times be target starved, making obtaining high observing efficiency more challenging. This can be seen in Figure \ref{fig:sac_eff}.

\subsubsection{Classical Strategy}
We follow a similar procedure to the large aperture strategy in Section \ref{sec:lat3}. Because the fields defined  extend to $10\degree$ declination, observations are limited to no more than $57\degree$ elevation, as seen in Figure \ref{fig:decaccess}. Since the telescope hardware must observe at or above $50\degree$ due to cryogenic constraints, we choose the observing elevations $50\degree$, $53.5\degree$, and $57\degree$.

Since, unlike the LAT strategy, larger regions are not split into northern and southern fields at the same right ascension, we can neglect the north/south split. Observing the three chosen elevations (one per day) either rising or setting make a total of six strategies. Complementary strategy pairs for Sun and Moon avoidance are listed in table \ref{tab:sat_complements}.

\begin{table}
\begin{center}
\begin{tabular}{| l | r |}
\hline
Strategy & Complement \\
\hline
Rising, $50\degree$ elevation & Setting, $57\degree$ elevation \\
Rising, $53.5\degree$ elevation & Setting, $53.5\degree$ elevation \\
Rising, $57\degree$ elevation & Setting, $50\degree$ elevation \\
\hline
\end{tabular}
\end{center}
\caption{
\label{tab:sat_complements}
Complement pairs for the small aperture classical avoidance strategy. Each line represents one pair.}
\end{table}

Unlike the large aperture strategy, there are large gaps in the local sidereal time that these fields can be observed if they are observed only rising or only setting on a given day. This would lead to significantly decreased observing efficiency. To counter this, gaps in the strategy are filled with the strategy's complement even before Sun or Moon avoidance are applied. This effectively observes fields on the opposite end of the sky at a lower priority. If this would cause the Sun or Moon to move within the avoidance radius, this observation will be removed as described in Section \ref{sec:avoidance}.

Finally, as with the large aperture strategy, scans less than 10 minutes in length are cut and each scan is made to have at least 60 seconds of idle time at its beginning to simulate telescope slewing. This results in a final observing efficiency of 64.6\%. It should be noted that this is significantly lower than the large aperture efficiency due to geometric constraints and a smaller observed sky area.

\subsubsection{Opportunistic Strategy}

\begin{figure}
\begin{center}
\includegraphics[width=\textwidth]{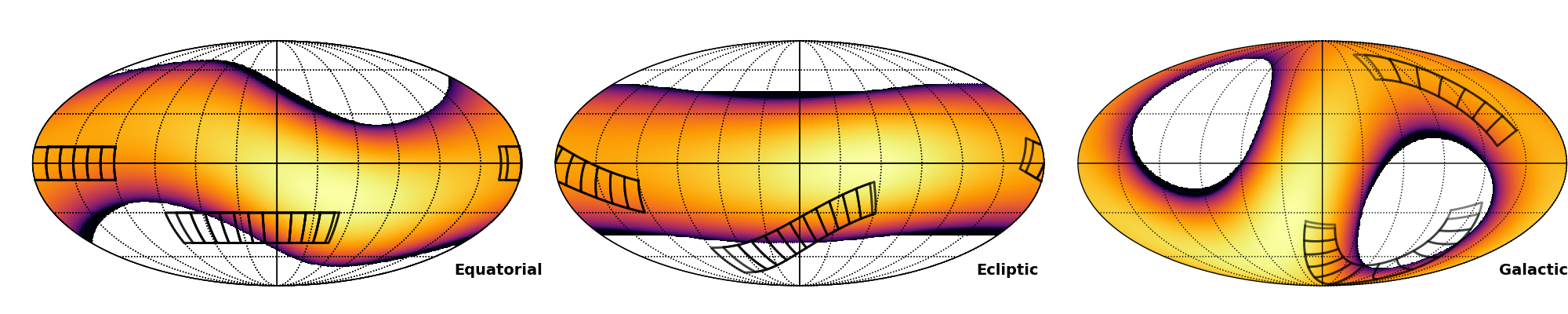}
\end{center}
\caption{
	\label{fig:sac_tiles}
	The tiling scheme for the opportunistic small area schedule. The colored
    band represents pixels that fall inside the $45^\circ$ avoidance region around the
    Sun or the Moon. The full patches form the outline of each patch and were given
    first priority while the $5^\circ\times10^\circ$ uniformly shaped tiles were only 
    targeted when the full patches could not be acquired.
}
\end{figure}

We created a small area opportunistic observing schedule by tiling the two patches in the RA direction into $10^\circ$ slices. To promote continuous scans of the patches we added the complete patches as first priority targets, instructing the scheduler to only consider tile scans when the full patches could not be scheduled, either due to to time or Sun avoidance constraints. This is done to improve the ability to separate ground synchronous modes from a real sky signal. The tiling is shown in Figure~\ref{fig:sac_tiles}. The resulting schedule has an observing efficiency of 61.1\% with most of the lost efficiency coming from August and September when the Sun covers significant parts of the northern patch.

\subsubsection{Improving the SAT observing efficiency}
Given the lower than desired survey strategy efficiencies described above, we plan to continue optimizing the field selection, elevation range, and Sun/Moon avoidance for the SAT telescopes.  For example, modest changes in the right ascension and declination, enabling lower elevation observations, and decreasing the Sun/Moon avoidance radii could significantly improve the observing efficiency.  Optimizing this parameter space should include consideration of instrument requirements, systematic mitigation, and the results of science forecasts, which we plan to pursue as we prepare for first light with the SATs.

\begin{figure}
\begin{center}
\begin{tabular}{l r}
\includegraphics[height=5cm]{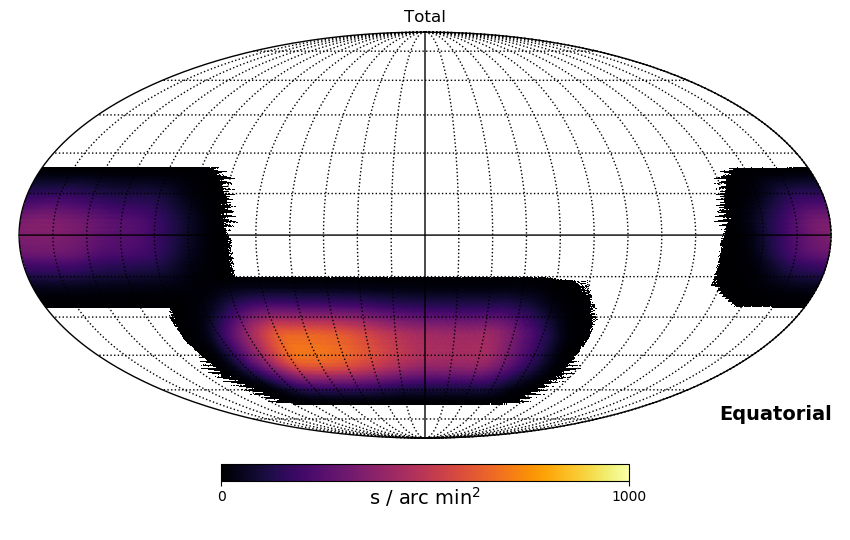} &
\includegraphics[height=5cm]{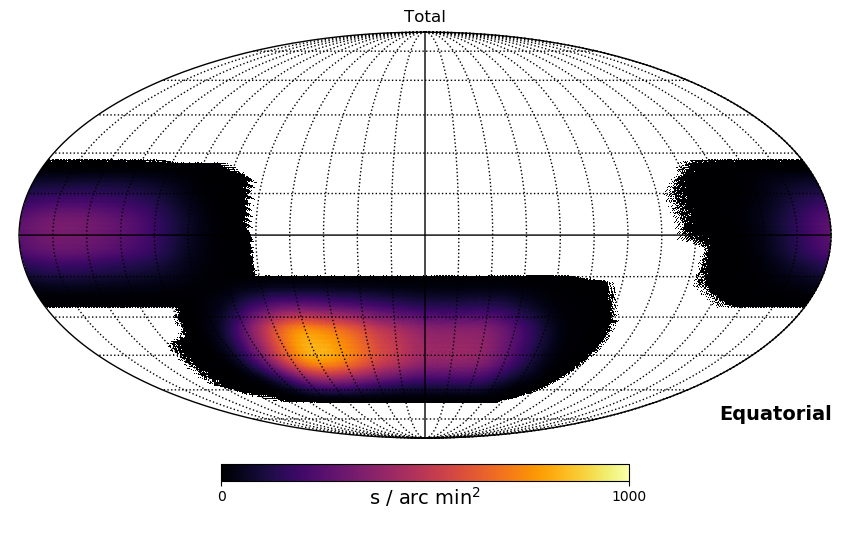} 
\end{tabular}
\end{center}
\caption{
	\label{fig:sac_comparison}
	Resulting integration depths from the classical (left) and opportunistic (right) schedules.
}
\end{figure}

\section{Comparison of Classical and Opportunistic Strategies}
\label{sec:comparison}

The ultimate goal of an observation strategy is to maximize the power of a given telescope to constrain cosmological parameters. The final parameter constraint will depend on many complex interrelated factors including instrumental noise and systematics, analysis pipeline implementation and observation strategy. As it is difficult to perfectly know and simulate all of these factors {\it a priori} it becomes important to define simplified figures of merit that can be easily estimated for a given strategy. We will discuss figures of merit that relate to the raw instrumental sensitivity including sky area and calendar efficiency as well as figures of merit that relate to the detection and suppression of systematic errors.

\begin{figure}
\begin{center}
\begin{tabular}{l r}
\includegraphics[height=6cm]{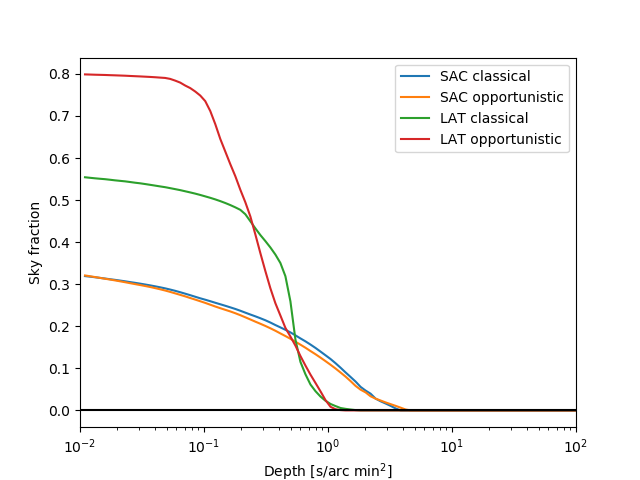} &
\includegraphics[height=5.5cm]{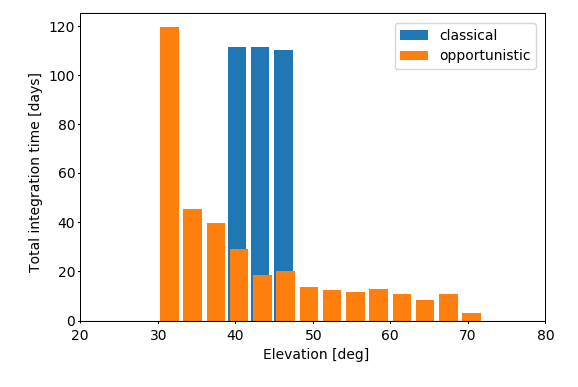} \\
\end{tabular}
\caption{
\label{fig:compare}
Left: The sky fraction deeper than a given depth for each strategy. Right: For both of the LAT strategies, the total amount of time spent observing at each elevation.
}
\end{center}
\end{figure}

\subsection{Optimizing Statistical Sensitivity}
\label{sec:fsky}

These figures of merit describe how efficiently a strategy allocates detector time within a survey area and how well the survey area is matched to the science objectives. The first figures of merit are the total and effective sky area covered. The total sky area is defined to be the total area for which there is any allocation of detector time. The effective sky area is defined to be\cite{2002ApJ...567....2H}:
$$
f_\textrm{sky,eff} = f_\textrm{sky,tot} \frac{w_2^2}{w_4}
$$
where $w^i$ is the $i\textrm{th}$ moment of the projected weight map. This term relates to the effective number of modes at a given $\ell$ and quantifies the trade-off between signal sample variance and noise variance.

Additionally, we define a figure of merit $\kappa$ that relates to the tapering of depth at the patch edges. Given a patch weight function $w(\hat{n})$, here approximated as a simple hit density map we can define a minimum depth $w_\textrm{eff}$ corresponding to the effective sky area:
$$
\iint_{w \geq w_\textrm{eff}} \textrm{d} \hat{n} = f_\textrm{sky,eff}
$$
We can then define a fraction of the total weight above this threshold:
$$
\kappa = \frac{\iint_{w \geq w_\textrm{eff}} w(\hat{n}) \textrm{d}\hat{n}}{\iint w(\hat{n}) \textrm{d}\hat{n}}
$$
In general a strategy with a higher value of $\kappa$ will have a more compact scan strategy that uses fewer detector hours on the low weight edges of a patch. It should be noted that this figure of merit is imperfect because some natural tapering at the patch edges is necessary to apodize maps for power spectrum estimation. There is an $\ell$-dependent optimal apodization window that approaches the inverse variance weight at high-$\ell$ and is as smooth as possible at low-$\ell$.\cite{2007PhRvD..76d3001S} In this sense a well designed low-$\ell$ strategy will incorporate some tapering at the map edges and will have $\kappa \leq 1$. The values of $f_\mathrm{sky,tot}$, $f_\mathrm{sky,eff}$, and $\kappa$ for each strategy are listed in Table \ref{tab:fsky}.




When selecting one of these strategies, there are several additional comparisons to consider. In the LAT opportunistic schedule it will be necessary to ensure adequate overlap between the tiles while including effects due to the finite time it takes the telescope to turn around. We will also need to study the  trade offs in power spectrum estimation between maps with more uniform versus more variable depths.

\begin{table}
\begin{center}
\begin{tabular}{| l | c | c | c |}
\hline
Strategy & $f_\mathrm{sky,tot}$ & $f_\mathrm{sky,eff}$ & $\kappa$ \\
\hline
LAT Classical & 0.575 & 0.246 & 0.637 \\
LAT Opportunistic & 0.804 & 0.264 & 0.611 \\
SAT Classical & 0.344 & 0.092 & 0.652 \\
SAT Opportunistic & 0.388 & 0.070 & 0.586 \\
\hline
\end{tabular}
\vspace{10pt}
\caption{
\label{tab:fsky}
Total and effective sky areas for each strategy, as well as $\kappa$. Note that the SAT strategies achieve significantly smaller sky areas than the LAT strategies. The effective sky areas and compactness $\kappa$ are similar between opportunistic and classical schedules.
}
\end{center}
\end{table}

\subsection{Optimizing Suppression of Systematics}
\label{subsec:systematics}

The scan strategy chosen for an experiment will also impact the ability to constrain and suppress instrumental systematics. A good scan strategy will scan the same portion of the sky at many different azimuths and elevations, and therefore many different detector angles on the sky. This will average down certain effects such as beam systematics and detector cross-talk. Additionally, the different coverage will allow for useful null tests that probe a wide range of possible systematics. The sensitivity of these null tests to different systematics is beyond the scope of this paper.

Additionally, for the LAT the different frequency tubes will not be exactly co-pointed meaning the coverage maps for each frequency will be naturally different, complicating efforts to separate the foreground and CMB components of the sky. A scan strategy can mitigate this effect by maximizing overlap between the different frequencies.

For the SAT there is an additional degree of freedom known as ``boresight rotation angle.'' This refers to the ability to rotate the refractive telescope around its boresight as a third axis of motion (in addition to azimuth and elevation). This has been deployed on a number of experiments to date including DASI, the Cosmic Background Imager, QUIET, QuAD and BICEP \cite{dasi1998,cbi2002,2009ApJ...692.1247P,2014ApJ...792...62B,2013ApJ...768....9B} as a powerful systematics mitigation strategy. This allows for additional detector angle rotation beyond what is already acheived by the scanning strategy. The SO LAT will also be capable of partial boresight rotations and partial instrument rotation. Both of these effects must be added to a final LAT survey strategy.


\FloatBarrier

\subsection{Detector Angle Rotation Figures of Merit}

One critical figure of merit of a scan strategy is the ability to suppress instrumental systematics by achieving varied detector angle coverage at each point on the map. This significantly mitigates systematics related to detector beam non-ideality such as differential pointing and ellipticity within a pair and crosstalk between neighboring pixels. 

It should be noted that such formalisms are written in the context of an experiment where the sky polarization signal is formed directly from polarization sensitive detectors such as the LAT, however varied detector angle coverage is also important in experiments where the sky polarization signal is modulated prior to the detectors (i.e. by a continuously rotating half wave plate) such as the SAT. This is because even detector angle coverage necessarily implies good cross linking, in which the scan paths at different times intersect significantly. This provides even $k$-space coverage on the sky and is necessary for good numerical behavior in maximum likelihood mapmaking algorithms.\cite{2002PhRvD..65b2003S,2010MNRAS.407.1387S,Poletti:2016xhi}

We will rely on a formalism developed in \cite{2017MNRAS.466..425W, ThMcBr} to quantify the evenness of detector angle coverage. Consider a differenced signal seen in a map pixel from two detector pairs {1,2} where the $\pi/4$ rotation in the second detector ensures sensitivity to both $Q$ and $U$,
\begin{equation}
    S^{d}_{\rm tot}(\psi) = h(\psi)S^{1}(\psi) + h(\psi - \pi/4)S^{2}(\psi).
    \label{eq:Sdtot}
\end{equation}
Here, $\psi_{j}$ is the angle of the $j$ th observation (hit) of the pixel, where
\begin{equation}
    h(\psi) = \frac{2\pi}{N_{\rm hits}}\sum_{j=1}^{N_{hits}}\delta(\psi - \psi_{j})
    \label{eq:h(psi)}
\end{equation}
characterizes the scan strategy in terms of the different crossing angles for each pixel. It is useful to expand this as a Fourier series to exploit the fact that certain beam non-idealities have an $n$-fold symmetry in map space. For example, a differential pointing signal is canceled by $\delta \psi = 180^\circ$ and a differential gain signal is canceled by $\delta \psi = 90^\circ$. We re-write $h(\psi)$ in the form:
\begin{equation}
    h(\psi) = \sum^{\infty}_{k=-\infty} \tilde{h}_{k} e^{ik\psi}
\end{equation}
where the $k$ denotes the spin of the $\tilde{h}_{k}$. 
These $\tilde{h}_{k}$ directly relate the amplitude of beam systematics in the polarization maps, where each $\tilde{h}_{k}$ corresponds to a beam systematic of spin $k$ (i.e. k=1 and k=3 corresponds to differential pointing and k=2 corresponds to differential gain). Taken to the  power spectrum level the $T\rightarrow P$ leakage from beam effects is suppressed by a factor of  $\langle \lvert\tilde{h}_{k}\rvert^{2} \rangle$:
\begin{equation}
    \langle \lvert\tilde{h}_{k}\rvert^{2} \rangle = \bigg\langle \bigg(\frac{1}{N_{hits}}\sum_{j=1}^{N_{hits}}\cos(k\psi_{j})\bigg)^{2} \bigg\rangle + \bigg\langle \bigg(\frac{1}{N_{hits}}\sum_{j=1}^{N_{hits}}\sin(k\psi_{j})\bigg)^{2} \bigg\rangle
    \label{eq:avgh_k}
\end{equation}
where the average is over all pixels observed by an experiment, $N_{hits}$ is the number of detector samples in a map pixel, and the $\psi_{j}$ is the detector angle of each hit.\cite{2009arXiv0906.1188B}

The $\langle \lvert\tilde{h}_{k}\rvert^{2} \rangle$ can be computed directly from the scan strategy and represent an overall susceptibility to beam non-idealities. Since it is difficult to optimize this concurrently with map coverage and observing efficiency we will treat this analysis in conjunction with optics simulations as a cross-check that beam systematics can be adequately controlled. A detailed study of this suppression of beam systematics will be presented in future work.

Additionally, the large aperture telescope can make use of a partial boresight rotation where the elevation of the beam can be swept over the zenith to the other side. This rotates the angle of each detector by $180 ^\circ$. If observations are evenly split between both configurations then effects such as differential pointing ($\tilde{h}_{1}$ and $\tilde{h}_{3}$) will be suppressed. This suggests that such partial boresight rotation can be a powerful tool to control beam systematics in the LAT and should be included in a scan strategy.


\FloatBarrier

\section{Future Work}

The scan strategies presented here represent proposals for the SO instruments based on heuristics to maximize the amount of telescope time that is spent on useful areas of the sky. There is still significant optimization work to be done in implementing full simulations. We quote approximate figures of merit for each of these strategies and discuss ways in which they may give a preliminary though incomplete picture of how the scan strategy impacts the final systematic and statistical uncertainties of measurements with the upcoming the SO telescopes and potentially next generation observations, such as CMB-S4.



\section{Acknowledgements}
This work was supported in part by a grant from the Simons Foundation (Award 457687,B.K.). MDN and JRS acknowledge support from NSF award AST-1454881. The technical scope of work was supported by the Laboratory Directed Research and Development Program of Lawrence Berkeley National Laboratory under U.S. Department of Energy Contract No. DE-AC02-05CH11231. This research used resources of the National Energy Research Scientific Computing Center (NERSC), a U.S. Department of Energy Office of Science User Facility operated under Contract No. DE-AC02-05CH11231.


\bibliographystyle{spiebib} 
\bibliography{main.bbl} 

\end{document}